\documentclass[twoside,preprint,fleqn,superscriptaddress,showkeys]{revtex4-2}
\usepackage{xcolor}    % Please do not change this line.
\usepackage{graphicx}  % Please do not change this line.
\usepackage{amssymb}   % Please do not change this line.
\usepackage{amsmath}   % Please do not change this line.
\usepackage{bm}        % Please do not change this line.
\usepackage{times}     % Please do not change this line.
\usepackage{fancyhdr}  % Please do not change this line.
\begin{document} % \setcounter{page}{1}  % Please do not change this line.

\title{Large anomalous Nernst effect in the ferromagnetic Fe$_3$Si polycrystal} 
\author{Yangming Wang} 
\affiliation{Department of Physics, University of Tokyo, Bunkyo-ku, Tokyo 113-0033, Japan} % first affiliation
\affiliation{Institute for Solid State Physics, University of Tokyo, Kashiwa, Chiba 277-8581, Japan}
\author{Susumu Minami}
\affiliation{Department of Physics, University of Tokyo, Bunkyo-ku, Tokyo 113-0033, Japan} % first affiliation

\author{Akito Sakai}
\affiliation{Department of Physics, University of Tokyo, Bunkyo-ku, Tokyo 113-0033, Japan} % first affiliation
\affiliation{Institute for Solid State Physics, University of Tokyo, Kashiwa, Chiba 277-8581, Japan}
\author{Taishi Chen}
\affiliation{Department of Physics, University of Tokyo, Bunkyo-ku, Tokyo 113-0033, Japan} % first affiliation
\affiliation{Institute for Solid State Physics, University of Tokyo, Kashiwa, Chiba 277-8581, Japan}
\affiliation{School of Physics, Southeast University, Nanjing, 211189, China}
\author{Zili Feng}
\affiliation{Department of Physics, University of Tokyo, Bunkyo-ku, Tokyo 113-0033, Japan} % first affiliation
\affiliation{Institute for Solid State Physics, University of Tokyo, Kashiwa, Chiba 277-8581, Japan}
\author{Daisuke Nishio-Hamane}
\affiliation{Institute for Solid State Physics, University of Tokyo, Kashiwa, Chiba 277-8581, Japan}
\author{Satoru Nakatsuji}
\email{satoru@phys.s.u-tokyo.ac.jp}
% Affiliation(s) of the third author 
\affiliation{Department of Physics, University of Tokyo, Bunkyo-ku, Tokyo 113-0033, Japan} % first affiliation
\affiliation{Institute for Solid State Physics, University of Tokyo, Kashiwa, Chiba 277-8581, Japan}
\affiliation{Institute for Quantum Matter and Department of Physics and Astronomy, Johns Hopkins University, Baltimore, Maryland 21218, USA} % second affiliation
\affiliation{CREST, Japan Science and Technology Agency (JST), 4-1-8 Honcho Kawaguchi, Saitama 332-0012, Japan} % first affiliation
\affiliation{Trans-scale Quantum Science Institute, University of Tokyo, Bunkyo-ku, Tokyo 113-8654, Japan}

%\date{\today} % Please do not change this line.

\begin{abstract}
The high-throughput calculation predicts that the Fe-based cubic ferromagnet Fe$_3$Si may exhibit a large anomalous Nernst effect (ANE). Here, we report our experimental observation of the large Nernst coefficient $S_{yx}\sim$2 $\mu$V/K and the transverse thermoelectric coefficient $-\alpha_{yx}$ $\sim$ 3 Am$^{-1}$K$^{-1}$ for Fe$_3$Si polycrystal at room temperature. The large $-\alpha_{yx}$ indicates that the large ANE originates from the intrinsic Berry curvature mechanism.
The high Curie temperature of 840 K and the most abundant raw elements of Fe and Si make Fe$_3$Si a competitive candidate for Nernst thermoelectric generations.
\end{abstract}


\maketitle  % Please do not change this line.
\thispagestyle{fancy}  % Please do not change this line.

\section{INTRODUCTION}

The thermoelectric (TE) effect, converting the heat current into electric energy directly, has a great potential for energy harvesting and heat flow sensors for advanced Internet of Things (IoT) society \cite{zhang2022micro,jaziri2020comprehensive,bell2008cooling}.
Ferromagnets can induce transverse thermoelectric voltages, so-called anomalous Nernst effect (ANE), which appears perpendicular to heat flow and magnetization. 
Recently, the ANE has attracted wide attention owing to its unique advantages for large-area and flexible thermoelectric devices \cite{nakatsuji2022topological,uchida2021transverse}. 
Thanks to recent developments in topological physics, giant ANE and anomalous Hall effect (AHE) enhanced by the large Berry curvature have been discovered in various ferromagnets and even antiferromagnets \cite{ikhlas2017large,sakai2020iron,sakai2018giant,guin2019zero,pan2022giant}. One of the most attractive candidates is Fe$_3X$ ($X$ = Ga, Al) where the Nernst coefficient $S_{yx}$ reaches up to 6 and 4  $\mu$V/K at room temperature in Fe$_3$Ga and Fe$_3$Al, respectively \cite{sakai2020iron}. The theoretical analysis indicates that such a giant ANE originated from the topological nodal web structure around the Fermi energy $E_{\rm F}$.

Here, we focus on a sister compound, cubic $D$0$_3$ Fe$_3$Si (Fig.\ref{fig1}(a)) \cite{ma2013theoretical}. The large ANE in Fe$_3$Si is predicted by the high throughput calculation \cite{sakai2020iron}. The Curie temperature ($T_{\rm C}$) of Fe$_3$Si $\sim 840 $ K is higher than that for Fe$_3$Ga (720 K) and Fe$_3$Al (600 K) \cite{niculescu1976relating,kawamiya1972magnetic,shinohara1964effect}, which is beneficial for TE application at high temperature. Besides, silicon is the most abundant element in the earth's crust and widely used for industry. However, the experimental report of ANE for Fe$_3$Si is limited only for thin films and its value is small $|S_{yx}|<\sim 1$ $\mu$V/K \cite{hamada2021anomalous}.

% Secondly, silicon is the most abundant solid-state element in nature. 
% Gallium is expensive and not easily stored due to the low melting point. For aluminum, the high pollution in the electrolysis process is becoming a serious environmental problems. Then it would be significant to replace Ga and Al by Si in Fe$_3X$. 
% What's more, Fe-Si alloy has been used as magnetic steel for voltage transformers with history of nearly 100 years, so the large scale process for this material has been well established. 

In this paper, we report the temperature ($T$) and magnetic field ($B$) dependence of the electric and thermoelectric properties for the bulk polycrystalline Fe$_3$Si. We find a large room-temperature Nernst coefficient $S_{yx}\sim$ 2 $\mu$V/K, which is twice larger than the previous report using thin films \cite{hamada2021anomalous}. 
We also estimate the transverse thermoelectric coefficient $|\alpha_{yx}|$ and found a large room-temperature value $\sim$ 3 Am$^{-1}$K$^{-1}$, suggesting the dominant intrinsic contribution. The analysis of the band structure indicates dispersion-less flat bands on the $\Gamma$-X line might have significant contribution to the large ANE response.

% \red{Although the nodal web structure in Fe$_{3}$Si is too far to affect ANE, a flat band structure within G-X region may have significant contribution, which is deserved for the further investigation.}

%In conclusion, low-cost raw elements and high Curie temperature make this system great potential in designing transverse thermoelectric devices.

\section{Experiments and discussion}
Polycrystalline Fe$_3$Si samples were synthesized by the melt cooling method in a mono-arc furnace. As-grown samples were used for all characterization and measurements.
The powder X-ray diffraction (XRD) result shows the single phase of the $D$0$_3$ Fe$_3$Si with a lattice constant of 5.65 \AA. 
The scanning Electron Microscope-Energy Dispersive X-ray Spectrometry (SEM-EDX) method shows our Fe$_3$Si is stoichiometric within a few percent resolutions.
The bar-shaped samples were used for all the transport properties, including Hall and longitudinal resistivity ($\rho_{yx}$ and $\rho_{xx}$), Nernst and Seebeck coefficients ($S_{yx}$ and $S_{xx}$) in a physical properties measurement system (PPMS, Quantum Design) with a thermal transport option (TTO). 
To remove the longitudinal contributions, $\rho_{xx}$ and $S_{xx}$, the temperature dependence of $\rho_{yx}$ and $S_{xx}$ were evaluated by symmetrization of the data with the positive and negative field sweeps.
Magnetization was measured by a commercial magnetic properties measurement system (MPMS, Quantum Design) with the needle-like sample ($\sim$ 0.2 mg). 
For all measurements, no specific orientation was chosen since the poly-crystalline nature could guarantee isotropic transport properties.

The electronic structure of Fe$_3$Si was obtained by using the OpenMX code \cite{OpenMX}, where the exchange-correlation functional within the generalized gradient approximation and norm-conserving pseudopotentials were employed \cite{PhysRevB.47.6728}.
The spin-orbit coupling was induced by using total angular momentum-dependent pseudopotentials.
The wave functions were expanded by a linear combination of multiple pseudoatomic orbitals \cite{PhysRevB.64.073106}.
A set of pseudoatomic orbital basis was specified as Fe5.5-$s3p2d2f1$, Si7.0-$s3p3d2$, where the number after each element stands for the radial cutoff in the unit of Bohr and the integer after $s, p, d,$ and $f$ indicates the radial multiplicity of each angular momentum component.
The lattice constant was set to the experimental lattice constant of 5.65 \AA.
The cutoff energies for a charge density of 500 Ry and a $k$-point mesh of $36\times 36\times 36$ were used.

\begin{figure}
\includegraphics[width = 0.8\columnwidth]{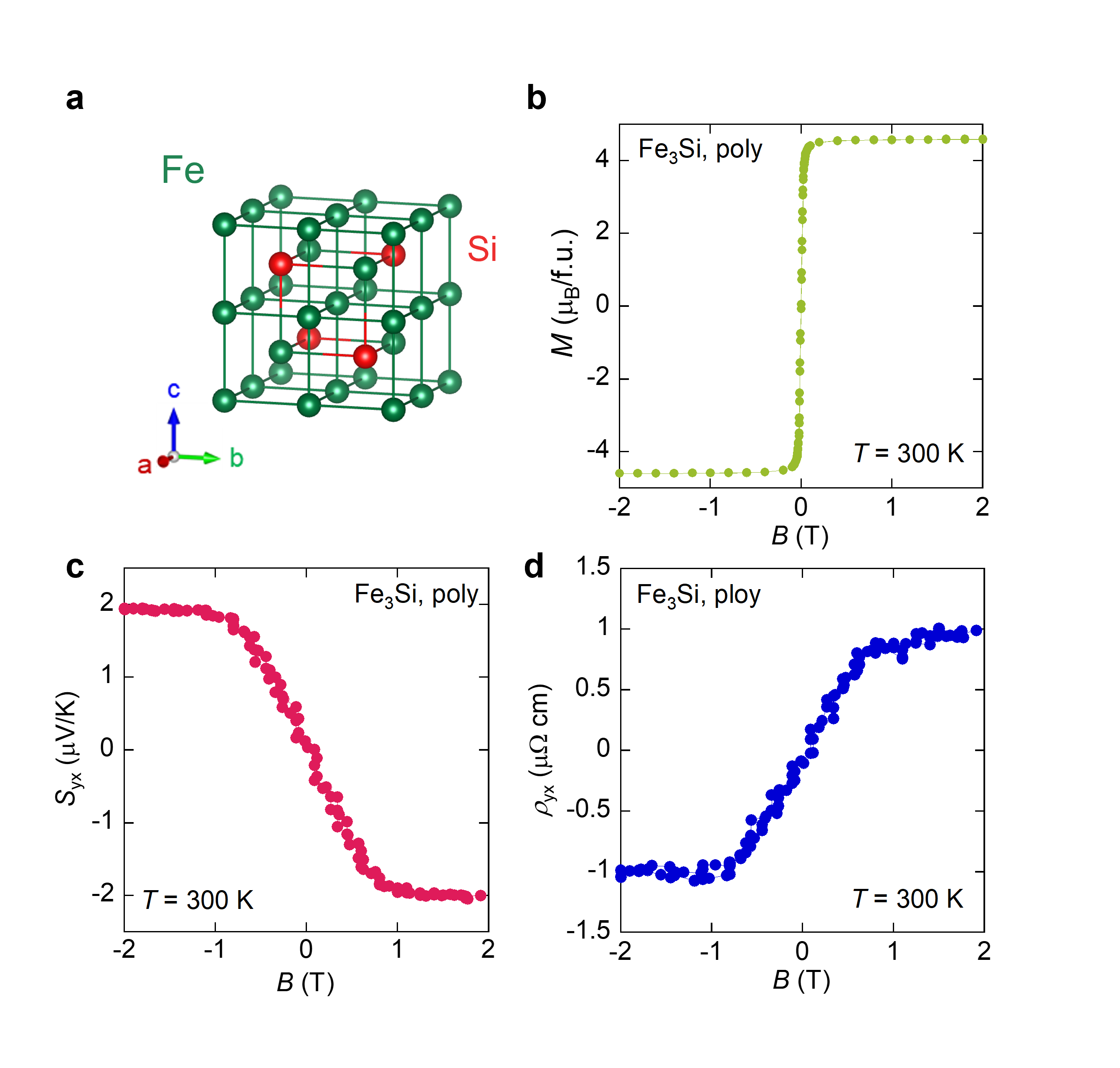} % Use any type of figure which is compatable with tex.
\caption{(Color online) (a) The ordered cubic $D$0$_3$ structure for Fe$_3$Si (Space group: $Fm\bar{3}m$) (b)-(d) The magnetic field dependence of the magnetization $M$ (b), the Nernst coefficient $S_{yx}$ (c) and the Hall resistivity $\rho_{yx}$ (d) for polycrystal Fe$_3$Si at 300 K. 
}
\label{fig1}
\end{figure}

Figure \ref{fig1}(b) shows the magnetic field dependence of magnetization $M$ at 300 K for Fe$_3$Si ferromagnet. The saturated magnetization $M_s$ is around 4.6 $\mu$$_B$/f.u at 300 K, which is comparable to the previous research ($M\sim$ 4.5 $\mu$$_B$/f.u) \cite{shinjo1963magnetic}. As predicted by the Slater-Pauling rule, $M_s$ for Fe$_3$Si is smaller than Fe$_3$Ga and Fe$_3$Al owing to the smaller number of valence electrons \cite{sakai2020iron}. 
Figure \ref{fig1} (c) and (d) show the $B$ dependence of the Nernst coefficient $S_{yx}$ and the Hall resistivity $\rho_{yx}$, respectively. Both $S_{yx}$ and $\rho_{yx}$ saturate at $B \sim$ 0.9 T. The difference between the saturated magnetic field in magnetization ($M$) and transport properties ($S_{yx}$ and $\rho_{yx}$) originates from the demagnetization effect owing to the shape anisotropy. $S_{yx}$ reaches $\sim$ 2 $\mu$V/K, which is nearly 10 times larger than that of pure iron \cite{watzman2016magnon}. This value is also comparable to the recent topological magnets such as kagome metal Fe$_3$Sn, Fe$_3$Sn$_2$  and TbMn$_6$Sn$_6$ %\textcolor{red}{...(more examples?)} 
\cite{zhang2021topological, chen2022large,zhang2022exchange}, suggesting some topological feature in the band structure may also be important in Fe$_3$Si. %The sign of $S_{yx}$ is defined as negative since the $S_{yx}$ exhibits the negative value under the positive field. Empirically, the sign of $S_{yx}$ in Fe-based alloys is negative, opposite to pure Fe \cite{sakai2020iron}. $\rho_{yx}$ at 300 K is $\sim$ 1 $\mathrm{\mu}\Omega$~cm with the opposite sign to $S_{yx}$.  

\begin{figure}
\includegraphics[width = 0.8\columnwidth]{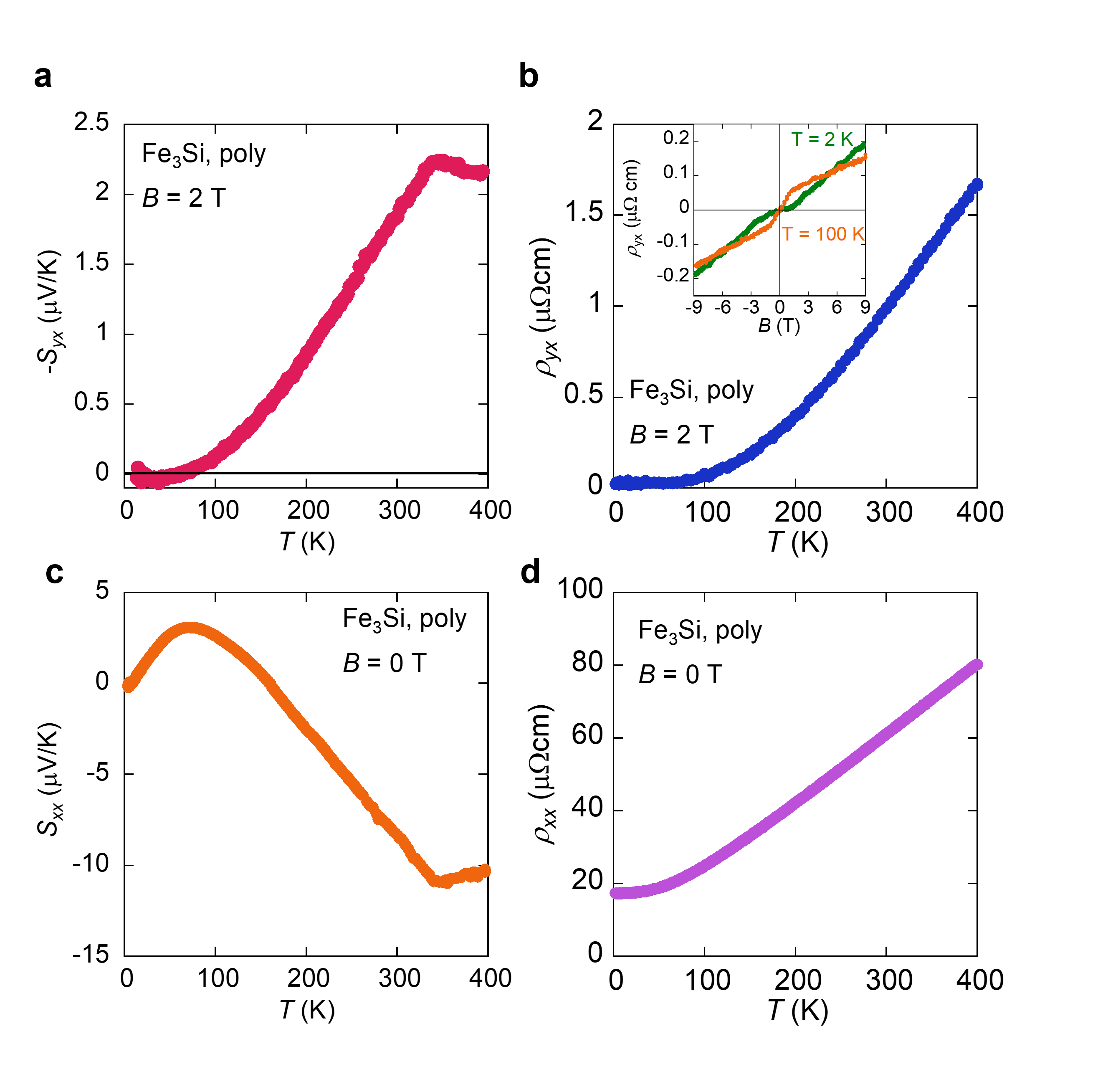} 
\caption{(Color online) 
(a, b) The temperature $T$ dependence of the Nernst coefficient -$S_{yx}$ (a) and Hall resistivity $\rho_{yx}$ (b) under the magnetic field of 2 T. The inset of Figure 2 (b) is the field dependence of $\rho_{yx}$ at 2 K and 100 K.
(c, d) The temperature $T$ dependence of the Seebeck coefficient $S_{xx}$ (c) and longitudinal resistivity $\rho_{xx}$ (d) under zero magnetic field. }
\label{fig2}
\end{figure}

To understand the mechanism of the large $S_{yx}$ in Fe$_3$Si, we also measure $T$ dependence of both transverse and longitudinal electric ($\rho_{yx}$ and $\rho_{xx}$) and thermoelectric properties ($S_{yx}$ and $S_{xx}$) as shown in Fig. \ref{fig2} . On cooling, $-S_{yx}$ peaks at $T\sim $ 340 K and then monotonically decreases down to the lowest temperature. $-S_{yx}$ becomes slightly negative below $\sim$ 70 K due to the carrier type change. Similarly, Seebeck coefficient $S_{xx}$ at zero fields also shows a peak around 340 K and monotonically decreases down to $\sim 70$ K accompanied by the sign change at $\sim 160$ K as shown in Fig.\ref{fig2} (c). The sign change of $S_{xx}$ indicates the change of the dominant carrier type from the electron at high $T$ to the hole low $T$.

On the other hand, $\rho_{yx}$ monotonically decreases down to the lowest temperature (Fig.\ref{fig2} (b)). Although $\rho_{yx}$ is governed by the AHE at high temperatures, clear $B$-linear ordinary Hall effect (OHE) contribution appears at low temperatures $T<\sim 100$ K and finally dominates at $T\sim 2$ K as shown in the inset of Fig.\ref{fig2} (b).
%Those contribution is driven by the Lorentz force rather than Berry curvature, thus it's non-intrinsic \cite{ye2018massive}. 
%At low temperature, the total value of $\rho_{yx}$ is tiny, thus the ratio of ONE is hard to ignore. For $S_{yx}$, the contribution from ordinary Nernst effect (ONE) should also exist. One possibly of tiny positive value of $S_{yx}$ at low temperature may be laso owing to ONE.
As shown in Fig.\ref{fig2} (d), $T$ dependence of $\rho_{xx}$ exhibits a typical metallic behavior, monotonically decreasing on cooling to the lowest temperature. Although the residual resistivity ratio (RRR) is similar to Fe$_{3}$Ga single crystals, the absolute value of $\rho_{xx}$ for Fe$_3$Si is only about 70$\%$ of that of Fe$_{3}$Ga  \cite{sakai2020iron}.

\begin{figure}
\includegraphics[width = 0.62\columnwidth]{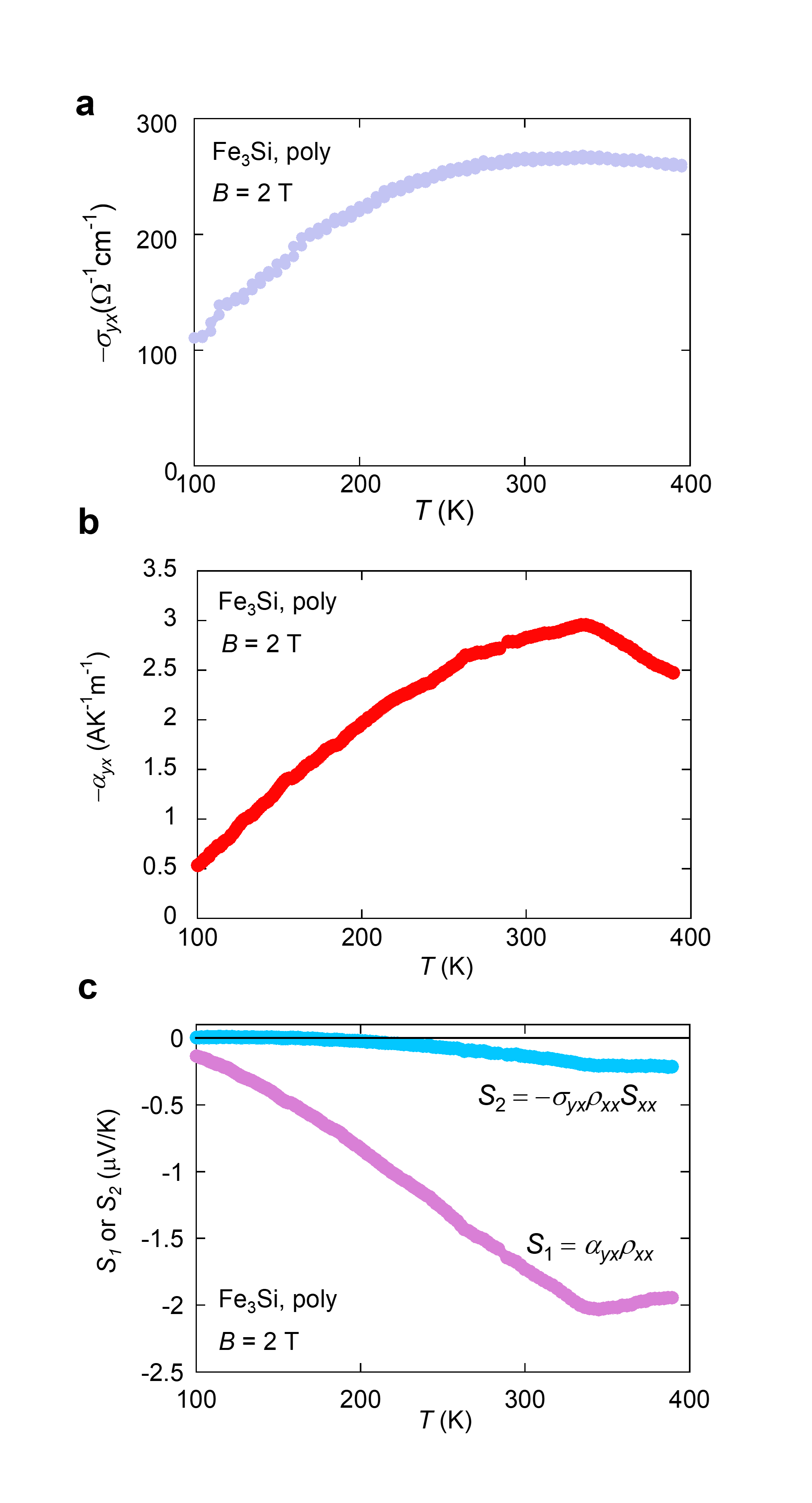} 
\caption{(Color online) The temperature $T$ dependence of the Hall conductivity
$-\sigma_{yx}$ (a), the transverse thermoelectric conductivity $-\alpha_{yx}$ (b), and two contributions $S_{1}$ and $S_{2}$ to anomalous Nernst effect (c) under the magnetic field of 2 T.
}\label{fig3}
\end{figure}

\begin{figure}
\includegraphics[width = 0.8\columnwidth]{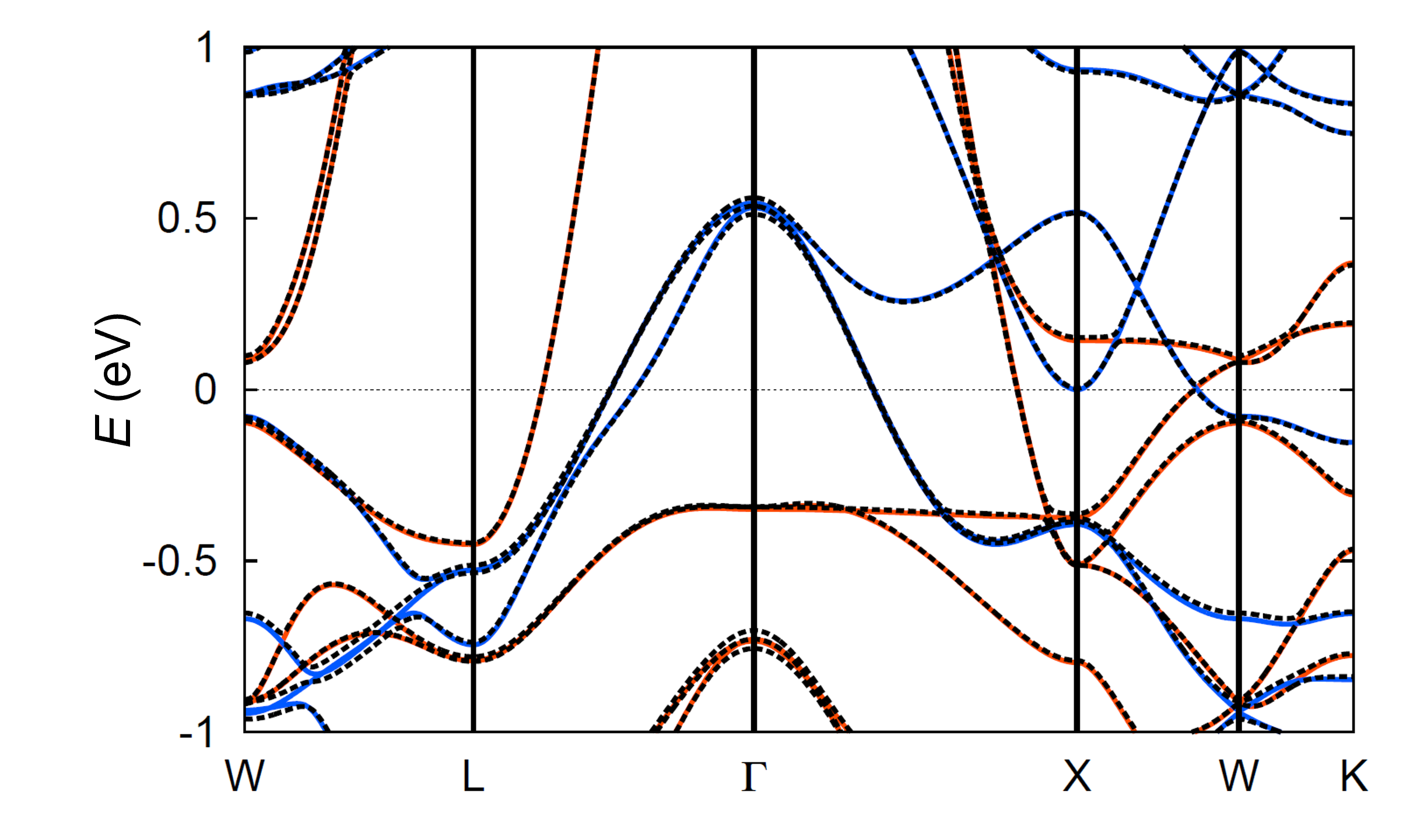} 
\caption{(Color online) Band structure around the Fermi energy for Fe$_3$Si obtained from first-principles calculations for the case of magnetization $M$ = 5.1 $\mu_B$/f.u. along [001]. The red and blue lines represent the majority and minority bands without SOC, respectively. The black dotted line corresponds to the bands with SOC.
}\label{fig4}
\end{figure}

To check the intrinsic contribution for ANE and AHE, we experimentally estimate the Hall conductivity $\sigma_{yx}$ and transverse thermoelectric conductivity $\alpha_{yx}$ based on the following formulas,
\begin{subequations}
\label{eq:one_A}
\begin{eqnarray}
\alpha_{yx} = \frac{1}{\rho_{xx}}[S_{yx}-\frac{\rho_{yx}}{\rho_{xx}}S_{xx}],
\\
\sigma_{yx}=-\rho_{yx}/\rho_{xx}^2.
\label{eq:one_B}
\end{eqnarray}
\end{subequations}
The obtained $-\sigma_{yx}$ and $-\alpha_{yx}$ are shown in Figs. \ref{fig3} (a) and \ref{fig3}(b), respectively. Here, we only show the data above $\sim 100$ K since we cannot easily separate the ordinal Hall/Nernst contribution at low temperatures as discussed above. As shown in Fig. \ref{fig3} (a), $-\sigma_{yx}$ shows a broad peak around 300 K and gradually decreases down to 100 K. On the other hand, $-\alpha_{yx}$ increases to 3 Am$^{-1}$K$^{-1}$ with a kink at $T\sim 340$ K and then monotonically decreases on cooling.

%In addition to the faint sign change at low temperature, $-\alpha_{yx}$ increases to 3 Am$^{-1}$K$^{-1}$ around 330 K and turns to decrease until 400 K. 
%As we discussed above, the abnormal $-\sigma_{yx}$ and $-\alpha_{yx}$ at low temperature may be contributed from the ordinary part OHE and ONE.
%which are hard to separate from temperature dependence dates. 
The equation (\ref{eq:one_A}a) can be rewritten as,
\begin{equation}
\label{eq:two}
    S_{yx}=\alpha_{yx}\rho_{xx} -\sigma_{yx}\rho_{xx} S_{xx}\equiv S_{1}+S_{2},
\end{equation}
The second term could be also expressed as $S_{2}$ =-tan$(\theta_{AHE}) S_{xx}$, where tan ($\theta_{AHE}$) is anomalous Hall angle. The first term $S_{1}$ represents the transverse voltage directly driven by the transverse thermoelectric coefficient $\alpha_{yx}$ while the second term $S_{2}$ can be regarded as the Hall effect of the current flow generated by the Seebeck effect.  
As shown in Figure \ref{fig3} (c), the contribution from $S_{2}$ is almost negligible around room temperature ($\sim$10$\%$ of $S_{1}$). This
indicates that the origin of the large ANE at room $T$ is indeed the large $\alpha_{yx}$.

%For the future material design, the relation between intrinsic transverse thermoelectric conductivity $|\alpha_{yx}|$ and Nernst voltage $S_{yx}$ was analysed. 
In fact, the room temperature $|\alpha_{yx}|$ for Fe$_3$Si is even larger than that for some topological materials with large ANE, such as Co$_2$MnGa [$|\alpha_{yx}|$ $\sim$ 2.7 Am$^{-1}$K$^{-1}$] and Fe$_3$Sn [$|\alpha_{yx}|$ $\sim$ 2.3 Am$^{-1}$K$^{-1}$] \cite{sakai2018giant,chen2022large}. However, the experimental $S_{yx}$ for Fe$_3$Si at 300 K is much smaller than Co$_2$MnGa [$S_{yx}$ $\sim$ 6.5 $\mu$V/K]. 
According to equation (\ref{eq:two}), this difference can be explained by two reasons. Firstly, the $\rho_{xx}$ for Fe$_3$Si is smaller than those topological semimetals \cite{sakai2018giant,chen2022large,sakai2020iron}. The Nernst voltage is the combination of the current flow induced by spin-orbital coupling (SOC) and the material resistance. 
Thus, the larger the longitudinal resistivity is, the larger ANE becomes if $|\alpha_{yx}|$ is the same. However, We note that the large resistivity suppresses the magnitude of the figure of merit $ ZT = \sigma S^2T/\kappa$, power factor $PF = \sigma S^2$ and the specific power generation capacity $\Gamma_{\mathrm{P}}=P_{\mathrm{max}}/\left(A\left(\Delta T\right)^{2}\right)$ in thermoelectric devices.
Therefore, semimetals with a resistivity of 100 $\sim$ 200 $\mathrm{\mu}\Omega$~cm at room temperature are more suitable for practical applications.
Secondly, the second term $S_{2}$ doesn't contribute too much to the total $S_{yx}$ in Fe$_3$Si. In Co$_2$MnGa, this contribution accounts for roughly 50 $\%$ of the total $S_{yx}$ because of the large Hall angle tan $(\theta_{AHE})$ \cite{sakai2018giant}. This phenomenon is also found in Co$_3$Sn$_2$S$_2$ \cite{guin2019zero,liu2018giant}. In order to utilize $S_{2}$ to improve the total $S_{yx}$, the tan$(\theta_{AHE})$ should be large $\sim$ 0.1, which is often discovered in topological ferromagnets with Weyl points \cite{li2020giant,armitage2018weyl}.

Figure \ref{fig4} shows the band structure around the Fermi energy for Fe$_3$Si obtained from first-principles calculation. 
Around the L point, nodal web structures similar to Fe$_3$Ga, composed of minority bands \cite{Minami_PRB_2020}, are found around $E=E_F - $0.6 eV. 
Unlike Fe$_3$Ga, the distance of the nodal web from $E_F$ is far to affect the transport properties. It indicates the origin of the large ANE for Fe$_3$Si might be different. In fact, we also found a flat band-like structure on the $\Gamma$-X line.
Since topological flat band structures could also be the source of giant ANE and AHE \cite{kato2022optical,roychowdhury2022large}, a further theoretical investigation is required to reveal the origin of the large transverse thermoelectric coefficient.

\section{CONCLUSIONS}
We have discovered a large ANE in the polycrystalline Fe$_3$Si, the sister compound of Fe$_3$Ga and Fe$_3$Al.
The one-step synthesis method and the low material cost make it the most promising material in the Fe$_3X$ system for future applications. In addition, for thin film applications, Fe$_3$Si could have better compatibility with other silicon-based electronic devices \cite{hines1976magnetization}.
% \red{The band structure analysis indicates the flat band within G-X region may have contribution to the large Berry curvature, it's natural to investigate the possible topological origins by analysing detailed band structure. }
The electronic band structure for Fe$_3$Si indicates that the dispersion-less band on the $\Gamma$-X line might induce a large Berry curvature instead of the nodal web structure like Fe$_3$Ga. However, the topological electronic structure as an origin of large ANE in Fe$_3$Si is still an open question.
For this purpose, systematic research based on single-crystal Fe$_3$Si is also expected in the future.

\begin{acknowledgments}
This work was partially supported by JST-Mirai Program (JPMJMI20A1), JST-CREST (JPMJCR18T3), New Energy and Industrial Technology Development Organization (NEDO), and JSPS-KAKENHI (JP19H00650, JP20K22479, JP21J22318, JP22K14587). 
The work at the Institute for Quantum Matter, an Energy Frontier Research Center was funded by DOE, Office of Science, Basic Energy Sciences under Award \# DE-SC0019331.
The computations in this research were partially carried out using the Fujitsu PRIMERGY CX400M1/CX2550M5 (Oakbridge-CX) in the Information Technology Center, The University of Tokyo.
And the use of the facilities of the Materials Design and Characterization Laboratory at the Institute for Solid State Physics, The University of Tokyo, is gratefully acknowledged.
\end{acknowledgments}

\end{document}